\documentclass[11pt,twoside,english]{article}
\usepackage[T1]{fontenc}
\usepackage[latin1]{inputenc}
\usepackage[letterpaper]{geometry}
\usepackage{color}
\usepackage{array}

\makeatletter

\providecommand{\tabularnewline}{\\}

\@ifundefined{definecolor}
 {\usepackage{color}}{}
\makeatother

\makeatother

\usepackage{babel}

\begin{document}

\title{Hierarchical quantum communication}

\maketitle
\begin{center}
Chitra Shukla$^{a}$ and Anirban Pathak$^{a,b,}$%
\footnote{email: anirban.pathak@gmail.com,

phone: +420 608650694%
} 
\par\end{center}

\begin{center}
$^{a}$Department of Physics and Materials Science Engineering, 
\par\end{center}

\begin{center}
Jaypee Institute of Information Technology, A-10, Sector-62, Noida,
UP-201307, India. 
\par\end{center}

\begin{center}
$^{b}$RCPTM, Joint Laboratory of Optics of Palacky University and
Institute of Physics of Academy of Science of the Czech Republic,
Faculty of Science, Palacky University, 17. listopadu 12, 771 46 Olomouc,
Czech Republic. 
\par\end{center}
\begin{abstract}
A general approach to study the hierarchical quantum information splitting
(HQIS) is proposed and the same is used to systematically investigate
the possibility of realizing HQIS using different classes of 4-qubit
entangled states that are not connected by SLOCC. Explicit examples
of HQIS using 4-qubit cluster state and 4-qubit $|\Omega\rangle$
state are provided. Further, the proposed HQIS scheme is generalized
to introduce two new aspects of hierarchical quantum communication.
To be precise, schemes of probabilistic hierarchical quantum information
splitting and hierarchical quantum secret sharing are obtained by
modifying the proposed HQIS scheme. A number of practical situations
where hierarchical quantum communication would be of use are also
presented.
\end{abstract}

\section{Introduction\label{sec:Introduction}}

In 1993 Bennett, Brassard, ${\rm Cr\acute{e}peau}$, Jozsa, Peres
and Wootters \cite{Bennet1993} proposed a scheme for quantum teleportation
using Bell state. This scheme drew considerable attention of the quantum
communication community as the teleportation has no classical analog.
Since the pioneering work of Bennett \emph{et al.} a large number
of teleportation schemes and their applications have been reported
\cite{Bennet1993,ijqi-qis,probalistic-1,probabilistic-2,probabilistic-3,probabilistic-4,qis-panigrahi,pankaj-qis-W,opt. comm,jphysb,ijtp,Hillery,Pati-RSP,Ba An -RSP}.
These teleportation schemes can be primarily classified into two broad
classes: a) Perfect teleportation schemes \cite{Bennet1993,ijqi-qis}
and b) Probabilistic teleportation schemes \cite{probalistic-1,probabilistic-2,probabilistic-3,probabilistic-4}.
By perfect teleportation we mean that the success rate of teleportation
is unity. This requires a maximally entangled quantum channel. However,
teleportation with unit fidelity is possible even when the quantum
channel is non-maximally entangled. In that case the success rate
of the receiver is not unity and the teleportation scheme is referred
to as the probabilistic teleportation scheme. Further, the teleportation
schemes are not limited to two-party teleportation (i.e., teleportation
between Alice and Bob). Many schemes of multi-party quantum teleportation
are also proposed in the last two decades (\cite{ijqi-qis,qis-panigrahi}
and references therein). Such multi-party teleportation schemes have
led to a set of interesting applications. Probably the most interesting
and most fundamental multi-party quantum teleportation scheme is the
controlled teleportation (CT) or the quantum information splitting
(QIS) scheme \cite{ijqi-qis,qis-panigrahi}. In this scheme, Alice
shares prior entanglement with Bob and at least one Charlie (supervisor).
Now nocloning theorem ensures that Alice cannot teleport copies of
the unknown quantum state (qubit) to both Charlie and Bob. Consequently,
if Alice succeeds to teleport an unknown quantum state to Charlie
and Bob then only one of them will be able to reconstruct the state
with the help of the other. Thus the quantum information is split
among Charlie and Bob and this is why such a scheme is called QIS
scheme. Now usually we assume that Bob recovers the unknown state
with the help of Charlie. Keeping this in mind, often Charlie is referred
to as the supervisor. Since Charlie can control the teleportation
protocol, this type of protocol is often referred to as CT protocol
too. Thus CT schemes are equivalent to QIS schemes. In recent past
CT/QIS schemes are reported using GHZ-like state \cite{ijqi-qis},
cluster state \cite{qis-panigrahi}, W state \cite{pankaj-qis-W},
etc. 

Till recent past all the studies on multi-party quantum teleportation
were restricted to symmetric quantum information splitting where all
the receivers were equally capable to recover the unknown quantum
state sent by the sender (Alice). Recently the concept of asymmetric
quantum information splitting is introduced by Wang \emph{et al.}
\cite{opt. comm,jphysb,ijtp}. In their scheme Boss (Alice) distributes
a quantum state among several agents who are spatially separated.
The agents are graded in accordance to their power to recover the
quantum state sent by Alice. A high power agent does not require the
help of all other agents to reconstruct the quantum state sent by
Alice, whereas a low power agent can reconstruct the secret state
iff all other agents cooperate with him. Thus there is a hierarchy
among the agents and consequently, the scheme is referred to as the
hierarchical quantum information splitting (HQIS) scheme. Wang \emph{et
al.} have recently reported HQIS using 4-qubit $|\chi\rangle$ state
\cite{opt. comm}, $t$-qubit graph state in general where $t\geq3$
\cite{jphysb}, and 6-qubit cluster state \cite{ijtp}. Extending
their idea here we propose a systematic and general procedure to investigate
the possibility of HQIS using an arbitrary ($n+1)$-qubit entangled
state and use our approach to explicitly show that the HQIS is possible
with 4-qubit $|\Omega\rangle$ state and 4-qubit cluster state. 

There exist several variants and applications of QIS. For example
one may think of probabilistic QIS or of Hillery, Buzek and Bertaiume's
\cite{Hillery} idea of quantum secret sharing (QSS) which may be
viewed as an application of QIS. So far all the schemes proposed for
QSS and probabilistic QIS are symmetric. This observation motivated
us to investigate the possibilities of introducing asymmetry (hierarchy)
among the receivers involved in a QSS or probabilistic QIS scheme.
The same is done here and that lead to the first ever protocols of
probabilistic HQIS and hierarchical QSS (HQSS). The proposed schemes
are interesting for several reasons. Primarily they are interesting
for their relevance in many practical situations. For example, we
may consider that Alice is president of a country and Diana is the
defense minister of that country. Bob and Charlie are defense secretary
and chief of  the armed forces of that country, respectively. Now
if the president wishes to permit the use of a nuclear weapon at a
suitable time then she distributes an information (say, a code which
is required to unlock the nuclear weapon) among the defense minister,
the defense secretary and the chief of the armed forces in such a
way that the minister can unlock the weapon if either the defense
secretary or the chief of the armed forces agrees and cooperates with
him. However, if the chief of the armed forces or the defense secretary
wants to unlock the weapon they would require the cooperation of each
other and that of the defense minister. Thus the defense minister
is more powerful than the chief of the armed forces and the defense
secretary, but even she is not powerful enough to unlock the weapon
alone. To realize this practically relevant scenario with unconditional
security we would require a scheme for HQSS. In what follows we will
see that HQSS using $|\Omega\rangle$ can be used to realize this
scenario. Here it would be apt to note that any attempt to implement
the above scenario using classical resources will not be unconditionally
secure. Another sector where HQSS is of everyday need is banking,
where a bank manager and/or cashier is usually more powerful than
the other users (office assistants and secretaries). However, even
the bank manager alone is not powerful enough to perform all the financial
operations related to his bank. For example, the password required
to unlock an ATM is always split into two or more pieces and the manager
alone cannot unlock it. Similarly, hierarchical secret sharing is
also essential for the smooth operation of the departmental stores.
The proposed schemes are also important because the existence of HQIS
automatically implies the existence of many related aspects of controlled
teleportation (e.g. controlled quantum information splitting, controlled
quantum secret sharing, controlled quantum state sharing etc.).

Remaining part of the paper is organized as follows. In Section \ref{sec:A-generalized-scheme}
we have described a generalized approach to investigate the possibility
of HQIS using $(n+1)$-qubit entangled states in general and as examples
we have explicitly shown that HQIS is possible using 4-qubit $|\Omega\rangle$
state and 4-qubit cluster state. In Section \ref{sec:PHQIS} we have
modified our scheme to obtain a scheme for probabilistic HQIS. The
HQIS scheme is further generalized in Section \ref{sec:HQSS} to yield
a scheme of HQSS. Finally, Section \ref{sec:Conclusions} is dedicated
for the conclusions.

\section{A generalized approach to perfect HQIS\label{sec:A-generalized-scheme}}

Let us start with a general $(n+1)$ qubit state of the form \begin{equation}
|\psi_{c}\rangle=\frac{1}{\sqrt{2}}[|0\rangle|\psi_{0}\rangle+|1\rangle|\psi_{1}\rangle],\label{eq:channel}\end{equation}
where $|\psi_{0}\rangle$ and $|\psi_{1}\rangle$ are arbitrary $n$
qubit state. Any arbitrary quantum state can be written in the above
form but for our purpose we would choose those states where $|\psi_{0}\rangle$
and $|\psi_{1}\rangle$ are orthogonal to each other%
\footnote{This choice would ensure that $\mathbf{|\psi_{c}\rangle}$ is an entangled
state. %
}. The subscript $c$ stands for channel. We consider that the first
qubit of $\mathbf{|\psi_{c}\rangle}$ is with Alice and the rest are
with $n$ agents, say ${\rm Bob_{1}},$ ${\rm Bob_{2}}$ etc. If we
work with $n=3$, we may call the agents as Bob, Charlie and Diana
and represent their respective qubits by the subscripts $B,\, C,\, D$.
Similarly, the qubit of Alice is indexed by the subscript $A.$ So
Alice shares $|\psi_{c}\rangle$ with $n$ agents and each agent has
one qubit.

Consider that Alice wishes to teleport (share) among her agents a
general one qubit state \begin{equation}
|\psi_{s}\rangle=\frac{1}{\sqrt{1+|\lambda|^{2}}}(|0\rangle+\lambda|1\rangle),\label{eq:chitra1}\end{equation}
 which represents an arbitrary qubit. So the combined state of Alice
and her agents is \begin{equation}
\begin{array}{lcl}
|\psi_{s}\rangle\otimes|\psi_{c}\rangle & = & \frac{1}{\sqrt{1+|\lambda|^{2}}}\left(|0\rangle+\lambda|1\rangle\right)\otimes\frac{1}{\sqrt{2}}\left[|0\rangle|\psi_{0}\rangle+|1\rangle|\psi_{1}\rangle\right]\\
 & = & \frac{1}{\sqrt{2(1+|\lambda|^{2})}}\left[|00\rangle|\psi_{0}\rangle+|01\rangle|\psi_{1}\rangle\right]+\frac{\lambda}{\sqrt{2(1+|\lambda|^{2})}}\left[|10\rangle|\psi_{0}\rangle+|11\rangle|\psi_{1}\rangle\right]\\
 & = & \frac{1}{2\sqrt{(1+|\lambda|^{2})}}\left[|\psi^{+}\rangle\left(|\psi_{0}\rangle+\lambda|\psi_{1}\rangle\right)+|\psi^{-}\rangle\left(|\psi_{0}\rangle-\lambda|\psi_{1}\rangle\right)\right.\\
 & + & \left.|\phi^{+}\rangle\left(|\psi_{1}\rangle+\lambda|\psi_{0}\rangle\right)+|\phi^{-}\rangle\left(|\psi_{1}\rangle-\lambda|\psi_{0}\rangle\right)\right].\end{array}\label{eq:chitra2}\end{equation}
Now Alice does a Bell measurement on the first 2 qubits. From (\ref{eq:chitra2})
we can see that after the Bell measurement of Alice the combined states
of all the $n$ agents reduces to $|\Psi^{\pm}\rangle=\frac{|\psi_{0}\rangle\pm\lambda|\psi_{1}\rangle}{\sqrt{1+|\lambda|^{2}}}$
or $|\Phi^{\pm}\rangle=\frac{|\psi_{1}\rangle\pm\lambda|\psi_{0}\rangle}{\sqrt{1+|\lambda|^{2}}}$.
To be precise, the complete relation between the outcome of the Bell
measurement of Alice and the combined state of the agents is given
in Table \ref{tab:Table-chitra1}, which is true in general. This
provides us the basic framework to investigate the possibilities of
HQIS in different quantum states. The remaining task is to appropriately
decompose the combined state of the agents so that one of them can
recover the unknown state and to find out the appropriate unitary
operation. In the following subsections we will consider two specific
cases and explicitly show that the above framework can be used to
establish that HQIS is possible with 4-qubit $|\Omega\rangle$ state
and 4-qubit cluster state.

\begin{table}
\begin{centering}
\begin{tabular}{|>{\centering}p{1.5in}|>{\centering}p{1.5in}|}
\hline 
Outcome of Alice's measurement  & Combined state of all agents after the measurement of Alice\tabularnewline
\hline 
$|\psi^{+}\rangle$  & $|\Psi^{+}\rangle=\frac{|\psi_{0}\rangle+\lambda|\psi_{1}\rangle}{\sqrt{1+|\lambda|^{2}}}$\tabularnewline
\hline 
$|\psi^{-}\rangle$  & $|\Psi^{-}\rangle=\frac{|\psi_{0}\rangle-\lambda|\psi_{1}\rangle}{\sqrt{1+|\lambda|^{2}}}$\tabularnewline
\hline 
$|\phi^{+}\rangle$  & $|\Phi^{+}\rangle=\frac{|\psi_{1}\rangle+\lambda|\psi_{0}\rangle}{\sqrt{1+|\lambda|^{2}}}$\tabularnewline
\hline 
$|\phi^{-}\rangle$  & $|\Phi^{-}\rangle=\frac{|\psi_{1}\rangle-\lambda|\psi_{0}\rangle}{\sqrt{1+|\lambda|^{2}}}$\tabularnewline
\hline
\end{tabular}
\par\end{centering}

\caption{\label{tab:Table-chitra1}Relation between Alice's measurement outcomes
and the combined states of the agents after the measurement of Alice.}
\end{table}

\subsection{Case I: \textmd{\normalsize $|\psi_{c}\rangle$ is 4-qubit $|\Omega\rangle$
state\label{sub:Case-I:-}}}

Let us assume that Alice has chosen 4-qubit $|\Omega\rangle$ state
\cite{Pradhan-Agarwal-Pati} as channel and kept the first photon
with her and has sent the second, third and fourth photons to Bob,
Charlie and Diana respectively. In that case \begin{equation}
\begin{array}{lcl}
|\psi_{c}\rangle=|\Omega\rangle_{ABCD} & = & \frac{1}{2}[|0000\rangle+|0110\rangle+|1001\rangle-|1111\rangle]_{ABCD}\\
 & = & \frac{1}{\sqrt{2}}[|0\rangle_{A}|\psi_{0}\rangle_{BCD}+|1\rangle_{A}|\psi_{1}\rangle_{BCD}],\end{array}\label{eq:4}\end{equation}
 where $|\psi_{0}\rangle_{BCD}=\frac{1}{\sqrt{2}}[|000\rangle+|110\rangle]$
and $|\psi_{1}\rangle_{BCD}=\frac{1}{\sqrt{2}}[|001\rangle-|111\rangle]$.

Now after Alice's Bell measurement on the first two qubits, the combined
state of Bob, Charlie and Diana collapses according to Table \ref{tab:Table-chitra1}.
If Alice's measurement outcome is $|\psi^{\pm}\rangle$ then the state
of the agents is

\begin{equation}
\begin{array}{lcl}
|\Psi^{\pm}\rangle_{BCD} & = & \frac{1}{\sqrt{1+|\lambda|^{2}}}[|\psi_{0}\rangle_{BCD}\pm\lambda|\psi_{1}\rangle_{BCD}]\\
 & = & \frac{1}{\sqrt{2(1+|\lambda|^{2})}}[|000\rangle+|110\rangle\pm\lambda(|001\rangle-|111\rangle)]_{BCD}.\end{array}\label{eq:chitra4}\end{equation}
 Similarly, if Alice obtains $|\phi^{\pm}\rangle$ then the state
of the agents is \begin{equation}
\begin{array}{lcl}
|\Phi^{\pm}\rangle_{BCD} & = & \frac{1}{\sqrt{1+|\lambda|^{2}}}[|\psi_{1}\rangle_{BCD}\pm\lambda|\psi_{0}\rangle_{BCD}]\\
 & = & \frac{1}{\sqrt{2(1+|\lambda|^{2})}}[|001\rangle-|111\rangle\pm\lambda(|000\rangle+|110\rangle)]_{BCD}.\end{array}\label{eq:chitra6}\end{equation}
 Now if the agents decide that Diana will recover the quantum state
sent by Alice, then we can decompose (\ref{eq:chitra4}) and (\ref{eq:chitra6})
as \begin{equation}
|\Psi^{\pm}\rangle_{BCD}=\frac{1}{\sqrt{2(1+|\lambda|^{2})}}[|00\rangle_{BC}(|0\rangle_{D}\pm\lambda|1\rangle_{D})+|11\rangle_{BC}(|0\rangle_{D}\mp\lambda|1\rangle_{D})]\label{eq:chitra5}\end{equation}
 and \begin{equation}
|\Phi^{\pm}\rangle_{BCD}=\frac{1}{\sqrt{2(1+|\lambda|^{2})}}[|00\rangle_{BC}(|1\rangle_{D}\pm\lambda|0\rangle_{D})-|11\rangle_{BC}(|1\rangle_{D}\mp\lambda|0\rangle_{D})].\label{eq:Chitra7}\end{equation}

\begin{table}
\begin{centering}
\begin{tabular}{|c|c|c|}
\hline 
Alice measurement outcome  & Bob, Charlie measurement outcome  & Diana's operation\tabularnewline
\hline 
$|\psi^{+}\rangle$  & $|00\rangle_{BC}$  & $I$\tabularnewline
\hline 
$|\psi^{+}\rangle$  & $|11\rangle_{BC}$  & $Z$\tabularnewline
\hline 
$|\psi^{-}\rangle$  & $|00\rangle_{BC}$  & $Z$\tabularnewline
\hline 
$|\psi^{-}\rangle$  & $|11\rangle_{BC}$  & $I$\tabularnewline
\hline 
$|\phi^{+}\rangle$  & $|00\rangle_{BC}$  & $X$\tabularnewline
\hline 
$|\phi^{+}\rangle$  & $|11\rangle_{BC}$  & $XZ$\tabularnewline
\hline 
$|\phi^{-}\rangle$  & $|00\rangle_{BC}$  & $XZ$\tabularnewline
\hline 
$|\phi^{-}\rangle$  & $|11\rangle_{BC}$  & $X$\tabularnewline
\hline
\end{tabular}
\par\end{centering}

\caption{\label{tab:Table2}Relation among the measurement outcomes of Alice,
Bob and Charlie and the unitary operations to be applied by Diana
when the initial state is an $|\Omega\rangle$ state and Diana reconstructs
the unknown state sent by Alice. Here the measurement outcomes of
Bob and Charlie are always same. So the communication from one of
them and Alice would be sufficient for Diana to reconstruct the unknown
state sent by Alice. }
\end{table}

Now form (\ref{eq:chitra5}) and (\ref{eq:Chitra7}) it is clear that
if Bob (Charlie) measures his (her) qubit in the computational basis
and sends the result to Diana, then Diana will be able to reconstruct
the state sent by Alice using appropriate unitary operators as shown
in Table \ref{tab:Table2}. For example, if Alice's measurement outcome
is $|\psi^{+}\rangle$ and that of Charlie is $|0\rangle$ then the
state of Diana is collapsed to $\frac{1}{\sqrt{1+|\lambda|^{2}}}(|0\rangle+\lambda|1\rangle)$,
so Diana needs to apply $I$. Thus Diana needs the help of Alice and
either Charlie or Bob to reconstruct the unknown state sent by Alice.
Now we would like to ask what happens if the agents decide that Bob
will reconstruct the state sent by Alice. From (\ref{eq:chitra4})
and (\ref{eq:chitra6}) it is clear that Charlie and Diana cannot
measure their qubits in the computational basis as that would collapse
the state of Bob to $|0\rangle$ or $|1\rangle$. Further, we note
that we can also decompose (\ref{eq:chitra4}) and (\ref{eq:chitra6})
as \begin{equation}
\begin{array}{lcl}
|\Psi^{\pm}\rangle_{BCD} & = & \frac{1}{2\sqrt{(1+|\lambda|^{2})}}\left[(|0\rangle_{B}\mp\lambda|1\rangle_{B})|\psi^{+}\rangle_{CD}+(|0\rangle_{B}\pm\lambda|1\rangle_{B})|\psi^{-}\rangle_{CD}\right.\\
 & + & \left.(|1\rangle_{B}\pm\lambda|0\rangle_{B})|\phi^{+}\rangle_{CD}-(|1\rangle_{B}\mp\lambda|0\rangle_{B})|\phi^{-}\rangle_{CD}\right],\end{array}\label{eq:chitra-8}\end{equation}

\begin{equation}
\begin{array}{lcl}
|\Phi^{\pm}\rangle_{BCD} & = & \frac{1}{2\sqrt{(1+|\lambda|^{2})}}\left[(|0\rangle_{B}\pm\lambda|1\rangle_{B})|\phi^{+}\rangle_{CD}+(|0\rangle_{B}\mp\lambda|1\rangle_{B})|\phi^{-}\rangle_{CD}\right.\\
 & - & \left.(|1\rangle_{B}\mp\lambda|0\rangle_{B})|\psi^{+}\rangle_{CD}+(|1\rangle_{B}\pm\lambda|0\rangle_{B})|\psi^{-}\rangle_{CD}\right].\end{array}\label{eq:chitra-9}\end{equation}

From (\ref{eq:chitra-8}) and (\ref{eq:chitra-9}) we can observe
that Bob can recover the arbitrary state $|\psi_{s}\rangle$ if a
Bell measurement can be done on the qubits of Charlie and Diana. This
can be fulfilled by either one of them (Charlie and Diana) communicating
a qubit to the other over an authenticated quantum channel, or both
performing a joint measurement (a nonlocal operation). The relation
among the measurement outcomes of Alice, Charlie and Diana and the
unitary operations to be applied by Bob to reconstruct the unknown
quantum state is explicitly provided in the Table \ref{tab:Table3}.
Here Charlie and Diana need to perform a joint measurement and consequently
Bob requires assistance of both of them and Alice to reconstruct the
unknown state sent by Alice. Thus Bob requires more information than
that required by Diana to reconstruct the unknown quantum state. Consequently,
Diana is more powerful than Bob%
\footnote{Charlie and Bob are equally powerful here. As the state of Charlie
and Bob are perfectly correlated, we have not explicitly described
the case where Charlie recovers the unknown quantum state.%
} and the scheme described here is a scheme of HQIS.

A special case of the above situation may be visualized as follows:
Alice sends Both C and D qubits to ${\rm Bob_{1}}$ and qubit B to
${\rm Bob_{2}}$. Now ${\rm Bob_{1}}$ can measure qubit C in computational
basis and apply unitary operation in accordance with Table \ref{tab:Table2}
 to reconstruct the state sent by Alice. To do so he does not require
any help of ${\rm Bob_{2}}$. However, ${\rm Bob_{2}}$ can reconstruct
the state iff ${\rm Bob_{1}}$ measures his qubits in Bell basis and
conveys the result to ${\rm Bob_{2}}$.

We may now note that the state with Bob in the first equation of (\ref{eq:chitra-8})
can be considered as \textit{quantum encrypted} with classical data
of 2 bits in the joint possession of Charlie and Diana, which is seen
manifestly as follows. \begin{equation}
|\Psi^{+}\rangle_{BCD}=\frac{1}{2}\left[(Z|\psi_{s}\rangle)_{B}|\psi^{+}\rangle_{CD}+(I|\psi_{s}\rangle)_{B}|\psi^{-}\rangle_{CD}+(X|\psi_{s}\rangle)_{B}|\phi^{+}\rangle_{CD}-(iY|\psi_{s}\rangle)_{B}|\phi^{-}\rangle_{CD}\right].\label{eq:chaitra-8}\end{equation}
 Without access to knowledge of the state with Charlie and Diana,
Bob's state is given by the reduced density operator: \begin{equation}
\frac{1}{4}\left(|\psi_{s}\rangle\langle\psi_{s}|+Z|\psi_{s}\rangle\langle\psi_{s}|Z+X|\psi_{s}\rangle\langle\psi_{s}|X+iY|\psi_{s}\rangle\langle\psi_{s}|iY\right)=\frac{I}{2},\end{equation}
implying that Bob gains no information without the cooperation of
Charlie and Diana. Analogous observations hold for the state $|\Psi^{-}\rangle$
in Eqn. (\ref{eq:chitra-8}) and states $|\Phi^{\pm}\rangle$ in Eqn.
(\ref{eq:chitra-9}).

\begin{table}
\begin{centering}
\begin{tabular}{|c|>{\centering}p{2in}|c|}
\hline 
Alice measurement outcome  &  Outcome of joint measurement of Charlie and Diana  & Bob's operation\tabularnewline
\hline 
$|\psi^{+}\rangle$  & $|\psi^{+}\rangle_{CD}$  & $Z$\tabularnewline
\hline 
$|\psi^{+}\rangle$  & $|\psi^{-}\rangle_{CD}$  & $I$\tabularnewline
\hline 
$|\psi^{+}\rangle$  & $|\phi^{+}\rangle_{CD}$  & $X$\tabularnewline
\hline 
$|\psi^{+}\rangle$  & $|\phi^{-}\rangle_{CD}$  & $XZ$\tabularnewline
\hline 
$|\psi^{-}\rangle$  & $|\psi^{+}\rangle_{CD}$  & $I$\tabularnewline
\hline 
$|\psi^{-}\rangle$  & $|\psi^{-}\rangle_{CD}$  & $Z$\tabularnewline
\hline 
$|\psi^{-}\rangle$  & $|\phi^{+}\rangle_{CD}$  & $XZ$\tabularnewline
\hline 
$|\psi^{-}\rangle$  & $|\phi^{-}\rangle_{CD}$  & $X$\tabularnewline
\hline 
$|\phi^{+}\rangle$  & $|\phi^{+}\rangle_{CD}$  & $I$\tabularnewline
\hline 
$|\phi^{+}\rangle$  & $|\phi^{-}\rangle_{CD}$  & $Z$\tabularnewline
\hline 
$|\phi^{+}\rangle$  & $|\psi^{+}\rangle_{CD}$  & $XZ$\tabularnewline
\hline 
$|\phi^{+}\rangle$  & $|\psi^{-}\rangle_{CD}$  & $X$\tabularnewline
\hline 
$|\phi^{-}\rangle$  & $|\phi^{+}\rangle_{CD}$  & $Z$\tabularnewline
\hline 
$|\phi^{-}\rangle$  & $|\phi^{-}\rangle_{CD}$  & $I$\tabularnewline
\hline 
$|\phi^{-}\rangle$  & $|\psi^{+}\rangle_{CD}$  & $X$\tabularnewline
\hline 
$|\phi^{-}\rangle$  & $|\psi^{-}\rangle_{CD}$  & $XZ$\tabularnewline
\hline
\end{tabular}
\par\end{centering}

\caption{\label{tab:Table3}Relation among the measurement outcomes of Alice,
Charlie and Diana and the unitary operations to be applied by Bob
when the initial state is an $|\Omega\rangle$ state and Bob reconstructs
the unknown state. Here Charlie and Diana needs to do a joint measurement
and consequently Bob requires assistance of both of them and Alice
to reconstruct the unknown state sent by Alice. }
\end{table}

\subsection{Case II: \textmd{\normalsize $|\psi_{c}\rangle$ is 4-qubit cluster
state }\textmd{($|C_{4}\rangle$)\label{sub:Case-II:-}}}

We assume that Alice has chosen 4-qubit cluster state ($|C_{4}\rangle$)
state as the channel and kept the first qubit with her and has sent
the second, third and fourth qubits to Bob, Charlie and Diana respectively.
In that case \begin{equation}
\begin{array}{lcl}
|\psi_{c}\rangle=|C_{4}\rangle_{ABCD} & = & \frac{1}{2}[|0000\rangle+|0011\rangle+|1100\rangle-|1111\rangle]_{ABCD}\\
 & = & \frac{1}{\sqrt{2}}[|0\rangle_{A}|\psi_{0}\rangle_{BCD}+|1\rangle_{A}|\psi_{1}\rangle_{BCD}],\end{array}\label{eq:4-1}\end{equation}
where $|\psi_{0}\rangle_{BCD}=\frac{1}{\sqrt{2}}[|000\rangle+|011\rangle]$
and $|\psi_{1}\rangle_{BCD}=\frac{1}{\sqrt{2}}[|100\rangle-|111\rangle]$.

Now after Alice's Bell measurement on the first two qubits, the combined
state of Bob, Charlie and Diana collapses according to Table \ref{tab:Table-chitra1}.
If Alice's measurement outcome is $|\psi^{\pm}\rangle$ then the combined
state of the agents is 

\begin{equation}
\begin{array}{lcl}
|\Psi^{\pm}\rangle_{BCD} & = & \frac{1}{\sqrt{1+|\lambda|^{2}}}[|\psi_{0}\rangle_{BCD}\pm\lambda|\psi_{1}\rangle_{BCD}]\\
 & = & \frac{1}{\sqrt{2(1+|\lambda|^{2})}}[|000\rangle+|011\rangle\pm\lambda(|100\rangle-|111\rangle)]_{BCD}.\end{array}\label{eq:chitra4-1}\end{equation}
Similarly, if Alice obtains $|\phi^{\pm}\rangle$ then the combined
state of the agents is \begin{equation}
\begin{array}{lcl}
|\Phi^{\pm}\rangle_{BCD} & = & \frac{1}{\sqrt{1+|\lambda|^{2}}}[|\psi_{1}\rangle_{BCD}\pm\lambda|\psi_{0}\rangle_{BCD}]\\
 & = & \frac{1}{\sqrt{2(1+|\lambda|^{2})}}[|100\rangle-|111\rangle\pm\lambda(|000\rangle+|011\rangle)]_{BCD}.\end{array}\label{eq:chitra6-1}\end{equation}
Now from (\ref{eq:chitra4}), (\ref{eq:chitra6}), (\ref{eq:chitra4-1})
and (\ref{eq:chitra6-1}) we can easily observe the following symmetry:
\[
\begin{array}{ccc}
\left.|\Psi^{\pm}\rangle_{BCD}\right|_{|\psi_{c}\rangle=|\Omega\rangle} & \equiv & \left.|\Psi^{\pm}\rangle_{DCB}\right|_{|\psi_{c}\rangle=|C_{4}\rangle},\\
\left.|\Phi^{\pm}\rangle_{BCD}\right|_{|\psi_{c}\rangle=|\Omega\rangle} & \equiv & \left.|\Phi^{\pm}\rangle_{DCB}\right|_{|\psi_{c}\rangle=|C_{4}\rangle}.\end{array}\]
Thus after the measurement of Alice the combined states of the agents
in this case (i.e., when $|\psi_{c}\rangle=|C_{4}\rangle$) is equivalent
to that in the previous case (i.e., when $|\psi_{c}\rangle=|\Omega\rangle$)
with the only difference that the role of Diana and Bob are now reversed.
Consequently, we obtain a HQIS scheme with $|\psi_{c}\rangle=|C_{4}\rangle_{ABCD}.$
However, here Bob is more powerful than Charlie and Diana. To be precise,
if Diana (Charlie) wishes to reconstruct the unknown quantum state,
then Bob and Charlie (Bob and Diana) have to perform a joint measurement
on their qubits. However, Bob will be able to reconstruct the state
using the measurement outcome of either Charlie or Diana. 

The framework used above to obtain these examples of HQIS is quite
general and may be used to investigate the possibility of perfect
HQIS using larger $n>3$ quantum states. However, we are not interested
in that. Rather we are interested to investigate the possibilities
of introducing hierarchy (asymmetry) in the other variants of QIS.
To be precise, in the following section we will investigate the possibility
of probabilistic HQIS and in the subsequent section we will introduce
hierarchical quantum secret sharing (HQSS).

\section{A scheme for probabilistic HQIS\label{sec:PHQIS}}

In the previous section we have described a scheme of perfect HQIS
using the maximally entangled state. Here we will introduce a scheme
of probabilistic HQIS. To do so we assume that Alice prepares and
shares with her agents a non-maximally entangled state of the form
$|\psi_{c^{\prime}}\rangle=a|0\rangle|\psi_{0}\rangle+b|1\rangle|\psi_{1}\rangle$,
where $|a|^{2}+|b|^{2}=1$ and $|a|\neq\frac{1}{\sqrt{2}}$. Now we
may follow the previous scheme of perfect teleportation to visualize
the difference. Here with the unknown state the initial state of the
system would become \begin{equation}
\begin{array}{lcl}
|\psi_{s}\rangle\otimes|\psi_{c^{\prime}}\rangle & = & \frac{1}{\sqrt{1+|\lambda|^{2}}}\left(|0\rangle+\lambda|1\rangle\right)\otimes\left[a|0\rangle|\psi_{0}\rangle+b|1\rangle|\psi_{1}\rangle\right]\\
 & = & \frac{1}{\sqrt{(1+|\lambda|^{2})}}\left[a|00\rangle|\psi_{0}\rangle+b|01\rangle|\psi_{1}\rangle\right]+\frac{\lambda}{\sqrt{(1+|\lambda|^{2})}}\left[a|10\rangle|\psi_{0}\rangle+b|11\rangle|\psi_{1}\rangle\right]\\
 & = & \frac{1}{\sqrt{2(1+|\lambda|^{2})}}\left[|\psi^{+}\rangle\left(a|\psi_{0}\rangle+\lambda b|\psi_{1}\rangle\right)+|\psi^{-}\rangle\left(a|\psi_{0}\rangle-\lambda b|\psi_{1}\rangle\right)\right.\\
 & + & \left.|\phi^{+}\rangle\left(b|\psi_{1}\rangle+\lambda a|\psi_{0}\rangle\right)+|\phi^{-}\rangle\left(b|\psi_{1}\rangle-\lambda a|\psi_{0}\rangle\right)\right].\end{array}\label{eq:chitra2-1}\end{equation}
Now Alice does a Bell measurement on the first 2 qubits. Therefore,
after the Bell measurement of Alice the combined state of all the
agents reduces to $|\Psi^{\prime\pm}\rangle=\frac{a|\psi_{0}\rangle\pm\lambda b|\psi_{1}\rangle}{\sqrt{|a|^{2}+|b\lambda|^{2}}}$
or $|\Phi^{\prime\pm}\rangle=\frac{b|\psi_{1}\rangle\pm\lambda a|\psi_{0}\rangle}{\sqrt{|b|^{2}+|a\lambda|^{2}}}$.
To be precise, the relation between the outcome of the Bell measurement
of Alice and the combined state of the agents are given in Table \ref{tab:Table-chitra1-1},
which is true in general. But to prove a specific example of probabilistic
HQIS let us consider that $|\psi_{c^{\prime}}\rangle$ is a non-maximally
entangled state of $|\Omega\rangle$ type and is described as \begin{equation}
\begin{array}{lcl}
|\psi_{c^{\prime}}\rangle=|\Omega^{\prime}\rangle_{ABCD} & = & [a|0000\rangle+a|0110\rangle+b|1001\rangle-b|1111\rangle]_{ABCD}\\
 & = & [a|0\rangle_{A}|\psi_{0}\rangle_{BCD}+b|1\rangle_{A}|\psi_{1}\rangle_{BCD}],\end{array}\label{eq:4-2}\end{equation}
where $|\psi_{0}\rangle_{BCD}=\frac{1}{\sqrt{2}}[|000\rangle+|110\rangle]$,
$|\psi_{1}\rangle_{BCD}=\frac{1}{\sqrt{2}}[|001\rangle-|111\rangle]$,
$|a|^{2}+|b|^{2}=1$ and $|a|\neq\frac{1}{\sqrt{2}}$ as before. 

\begin{table}
\begin{centering}
\begin{tabular}{|>{\centering}p{1.5in}|>{\centering}p{1.5in}|}
\hline 
Outcome of Alice's measurement  & Combined state of all agents after the measurement of Alice\tabularnewline
\hline 
$|\psi^{+}\rangle$  & $|\Psi^{\prime+}\rangle=\frac{a|\psi_{0}\rangle+\lambda b|\psi_{1}\rangle}{\sqrt{|a|^{2}+|b\lambda|^{2}}}$\tabularnewline
\hline 
$|\psi^{-}\rangle$  & $|\Psi^{\prime-}\rangle=\frac{a|\psi_{0}\rangle-\lambda b|\psi_{1}\rangle}{\sqrt{|a|^{2}+|b\lambda|^{2}}}$\tabularnewline
\hline 
$|\phi^{+}\rangle$  & $|\Phi^{\prime+}\rangle=\frac{b|\psi_{1}\rangle+\lambda a|\psi_{0}\rangle}{\sqrt{|b|^{2}+|a\lambda|^{2}}}$\tabularnewline
\hline 
$|\phi^{-}\rangle$  & $|\Phi^{\prime-}\rangle=\frac{b|\psi_{1}\rangle-\lambda a|\psi_{0}\rangle}{\sqrt{|b|^{2}+|a\lambda|^{2}}}$\tabularnewline
\hline
\end{tabular}
\par\end{centering}

\caption{\label{tab:Table-chitra1-1}Relation between Alice's measurement outcomes
and the combined states of the agents after the measurement of Alice.
Here the quantum channel is non-maximally entangled.}
\end{table}

Now in this particular case after Alice's Bell measurement on the
first two qubits, the combined state of Bob, Charlie and Diana collapses
according to Table \ref{tab:Table-chitra1-1}. If Alice's measurement
outcome is $|\psi^{\pm}\rangle$ then the combined state of the agents
is

\begin{equation}
\begin{array}{lcl}
|\Psi^{\prime\pm}\rangle_{BCD} & = & \frac{1}{\sqrt{|a|^{2}+|b\lambda|^{2}}}[a|\psi_{0}\rangle_{BCD}\pm\lambda b|\psi_{1}\rangle_{BCD}]\\
 & = & \frac{1}{\sqrt{|a|^{2}+|b\lambda|^{2}}}[a(|000\rangle+|110\rangle)\pm\lambda b(|001\rangle-|111\rangle)]_{BCD}\end{array}\label{eq:chitra4-2}\end{equation}
 Similarly, if Alice obtains $|\phi^{\pm}\rangle$ then the state
of the agents is \begin{equation}
\begin{array}{lcl}
|\Phi^{\prime\pm}\rangle_{BCD} & = & \frac{1}{\sqrt{|b|^{2}+|a\lambda|^{2}}}[b|\psi_{1}\rangle_{BCD}\pm\lambda a|\psi_{0}\rangle_{BCD}]\\
 & = & \frac{1}{\sqrt{|b|^{2}+|a\lambda|^{2}}}[b\left(|001\rangle-|111\rangle\right)\pm\lambda a(|000\rangle+|110\rangle)]_{BCD}\end{array}\label{eq:chitra6-2}\end{equation}
 Now if the agents decide that Diana will recover the quantum state,
then we can decompose (\ref{eq:chitra4-2}) and (\ref{eq:chitra6-2})
as \begin{equation}
|\Psi^{\prime\pm}\rangle_{BCD}=\frac{1}{\sqrt{|a|^{2}+|b\lambda|^{2}}}[|00\rangle_{BC}(a|0\rangle_{D}\pm\lambda b|1\rangle_{D})+|11\rangle_{BC}(a|0\rangle_{D}\mp\lambda b|1\rangle_{D})]\label{eq:chitra5-1}\end{equation}
 and \begin{equation}
|\Phi^{\prime\pm}\rangle_{BCD}=\frac{1}{\sqrt{|b|^{2}+|a\lambda|^{2}}}[|00\rangle_{BC}(b|1\rangle_{D}\pm\lambda a|0\rangle_{D})-|11\rangle_{BC}(b|1\rangle_{D}\mp\lambda a|0\rangle_{D})].\label{eq:Chitra7-1}\end{equation}
Now Bob and Charlie measure their respective qubit in computational
basis. Alice has already measured her qubits in Bell basis. Up to
this point this protocol is similar to the previous protocol of perfect
HQIS. But now Diana will not be able to obtain the unknown state just
by following the previous protocol of perfect HQIS. For example, if
Bob informs that his measurement outcome is $|0\rangle$ and Alice
informs that her measurement outcome is $|\psi^{+}\rangle$. Then
Diana's state is $|\psi\rangle_{1}=\frac{a|0\rangle_{D}+\lambda b|1\rangle_{D}}{\sqrt{|a|^{2}+|b\lambda|^{2}}}\neq\frac{1}{\sqrt{1+|\lambda|^{2}}}\left(|0\rangle+\lambda|1\rangle\right)_{D}.$
Thus Diana cannot follow previous protocol and apply $I$ to reconstruct
the unknown state sent by Alice. In fact, Diana cannot construct a
single qubit unitary operation to map $\frac{a|0\rangle_{D}+\lambda b|1\rangle_{D}}{\sqrt{|a|^{2}+|b\lambda|^{2}}}$
to $\frac{1}{\sqrt{1+|\lambda|^{2}}}\left(|0\rangle+\lambda|1\rangle\right)_{D}$
without the knowledge of $\lambda$. Therefore, Diana has to change
her strategy as follows. 

Diana prepares an ancilla qubit in $|0\rangle_{Auxi}$ and applies
following unitary operation on her qubits (i.e., on the combined system
of her existing qubit and ancilla): \begin{equation}
U=\left(\begin{array}{cccc}
\frac{b}{a} & \sqrt{1-\frac{b^{2}}{a^{2}}} & 0 & 0\\
0 & 0 & 0 & -1\\
0 & 0 & 1 & 0\\
\sqrt{1-\frac{b^{2}}{a^{2}}} & -\frac{b}{a} & 0 & 0\end{array}\right).\label{eq:U-probabilistic-teleportation}\end{equation}
As $a$ and $b$ are known, construction of $U$ is allowed. Now in
the specific case considered above (where Alice's measurement outcome
is $|\psi^{+}\rangle$ and that of Bob is $|0\rangle)$, Diana applies
$U$ on her product state \[
|\psi\rangle_{2}=|\psi\rangle_{1}|0\rangle_{Auxi}=\left(\frac{a|0\rangle_{D}+\lambda b|1\rangle_{D}}{\sqrt{|a|^{2}+|b\lambda|^{2}}}\right)|0\rangle_{Auxi}=\frac{1}{\sqrt{|a|^{2}+|b\lambda|^{2}}}\left(\begin{array}{c}
a\\
0\\
b\lambda\\
0\end{array}\right)\]
and obtains \[
\begin{array}{lcl}
U|\psi\rangle_{2} & = & \frac{1}{\sqrt{|a|^{2}+|b\lambda|^{2}}}\left(\begin{array}{c}
b\\
0\\
b\lambda\\
\sqrt{1-\frac{b^{2}}{a^{2}}}a\end{array}\right)\\
 & = & \frac{1}{\sqrt{|a|^{2}+|b\lambda|^{2}}}\left(b(|0\rangle+\lambda|1\rangle)|0\rangle+\sqrt{a^{2}-b^{2}}|1\rangle|1\rangle\right).\end{array}\]
Now Diana measures the last qubit (ancilla) in the computational basis.
If her measurement yields $|0\rangle$ then she obtains unknown state
with unit fidelity but if her measurement on ancilla yields $|1\rangle$
then the HQIS scheme fails. Similarly, we can check the other 15 possibilities.
The complete table that relates Alice's measurement outcomes, Bob's
(Charlie's) measurement outcomes, Diana's measurement outcomes and
unitary operations to be applied by Diana and Diana's conclusions
are given in the Table \ref{tab:Probabilistic-teleportation:-Relation}.

\begin{table}
\begin{centering}
\begin{tabular}{|>{\centering}p{0.6in}|>{\centering}p{0.6in}|>{\centering}p{0.6in}|>{\centering}p{0.8in}|>{\centering}p{0.8in}|>{\centering}p{0.8in}|}
\hline 
Alice's measurement outcome & Bob's measurement outcome & Diana's measurement 

outcome & Diana's state & Operation applied by Diana & Final state of Diana\tabularnewline
\hline 
$|\psi^{+}\rangle$ & $|0\rangle$ & $|0\rangle$ & $\frac{|0\rangle+\lambda|1\rangle}{\sqrt{1+|\lambda|^{2}}}$ & $I$ & $\frac{|0\rangle+\lambda|1\rangle}{\sqrt{1+|\lambda|^{2}}}$\tabularnewline
\hline 
$|\psi^{-}\rangle$ & $|0\rangle$ & $|0\rangle$ & $\frac{|0\rangle-\lambda|1\rangle}{\sqrt{1+|\lambda|^{2}}}$ & $Z$ & $\frac{|0\rangle+\lambda|1\rangle}{\sqrt{1+|\lambda|^{2}}}$\tabularnewline
\hline 
$|\phi^{+}\rangle$ & $|0\rangle$ & $|0\rangle$ & $\frac{\lambda|0\rangle+|1\rangle}{\sqrt{1+|\lambda|^{2}}}$ & $X$ & $\frac{|0\rangle+\lambda|1\rangle}{\sqrt{1+|\lambda|^{2}}}$\tabularnewline
\hline 
$|\phi^{-}\rangle$ & $|0\rangle$ & $|0\rangle$ & $-\frac{\left(\lambda|0\rangle-|1\rangle\right)}{\sqrt{1+|\lambda|^{2}}}$ & $iY$ & $\frac{|0\rangle+\lambda|1\rangle}{\sqrt{1+|\lambda|^{2}}}$\tabularnewline
\hline 
$|\psi^{+}\rangle$ & $|1\rangle$ & $|0\rangle$ & $\frac{|0\rangle-\lambda|1\rangle}{\sqrt{1+|\lambda|^{2}}}$ & $Z$ & $\frac{|0\rangle+\lambda|1\rangle}{\sqrt{1+|\lambda|^{2}}}$\tabularnewline
\hline 
$|\psi^{-}\rangle$ & $|1\rangle$ & $|0\rangle$ & $\frac{|0\rangle+\lambda|1\rangle}{\sqrt{1+|\lambda|^{2}}}$ & $I$ & $\frac{|0\rangle+\lambda|1\rangle}{\sqrt{1+|\lambda|^{2}}}$\tabularnewline
\hline 
$|\phi^{+}\rangle$ & $|1\rangle$ & $|0\rangle$ & $\frac{\lambda|0\rangle-|1\rangle}{\sqrt{1+|\lambda|^{2}}}$ & $iY$ & $\frac{|0\rangle+\lambda|1\rangle}{\sqrt{1+|\lambda|^{2}}}$\tabularnewline
\hline 
$|\phi^{-}\rangle$ & $|1\rangle$ & $|0\rangle$ & $-\frac{\left(\lambda|0\rangle+|1\rangle\right)}{\sqrt{1+|\lambda|^{2}}}$ & $X$ & $\frac{|0\rangle+\lambda|1\rangle}{\sqrt{1+|\lambda|^{2}}}$\tabularnewline
\hline
\end{tabular}
\par\end{centering}

\caption{\label{tab:Probabilistic-teleportation:-Relation}Probabilistic HQIS
(Diana reconstructs the unknown state): Relation among Alice's measurement
outcomes, Bob's (Charlie's) measurement outcomes, Diana's measurement
outcomes, unitary operations used by Diana and Diana's conclusions/final
state. In the remaining 8 cases where Diana's measurement outcome
is $|1\rangle$, her final state will be $|1\rangle$ and the HQIS
scheme will fail. Global phases are neglected in the last column. }
\end{table}

Now consider that Bob recovers the state. In that case we can decompose
(\ref{eq:chitra4-2}) and (\ref{eq:chitra6-2}) as \begin{equation}
\begin{array}{lcl}
|\Psi^{\prime\pm}\rangle_{BCD} & = & \frac{1}{\sqrt{2(|a|^{2}+|\lambda b|^{2})}}\left[(a|0\rangle_{B}\mp\lambda b|1\rangle_{B})|\psi^{+}\rangle_{CD}+(a|0\rangle_{B}\pm\lambda b|1\rangle_{B})|\psi^{-}\rangle_{CD}\right.\\
 & + & \left.(a|1\rangle_{B}\pm\lambda b|0\rangle_{B})|\phi^{+}\rangle_{CD}-(a|1\rangle_{B}\mp\lambda b|0\rangle_{B})|\phi^{-}\rangle_{CD}\right],\end{array}\label{eq:chitra-8-1}\end{equation}

\begin{equation}
\begin{array}{lcl}
|\Phi^{\prime\pm}\rangle_{BCD} & = & \frac{1}{\sqrt{2(|\lambda a|^{2}+|b|^{2})}}\left[(b|0\rangle_{B}\pm\lambda a|1\rangle_{B})|\phi^{+}\rangle_{CD}+(b|0\rangle_{B}\mp\lambda a|1\rangle_{B})|\phi^{-}\rangle_{CD}\right.\\
 & - & \left.(b|1\rangle_{B}\mp\lambda a|0\rangle_{B})|\psi^{+}\rangle_{CD}+(b|1\rangle_{B}\pm\lambda a|0\rangle_{B})|\psi^{-}\rangle_{CD}\right].\end{array}\label{eq:chitra-9-1}\end{equation}
Bob prepares an ancilla qubit in $|0\rangle_{Auxi}$ and applies the
2-qubit unitary operator $U$ described by (\ref{eq:U-probabilistic-teleportation})
on his qubits (i.e., on the combined system of his existing qubit
and ancilla) iff the measurement outcome of combined measurement of
Charlie and Diana is $|\psi^{\pm}\rangle$, otherwise he applies \[
U_{1}=U\left(X\otimes I\right)=\left(\begin{array}{cccc}
0 & 0 & \frac{b}{a} & \sqrt{1-\frac{b^{2}}{a^{2}}}\\
0 & -1 & 0 & 0\\
1 & 0 & 0 & 0\\
0 & 0 & \sqrt{1-\frac{b^{2}}{a^{2}}} & -\frac{b}{a}\end{array}\right)\]
on his qubits. To be precise, if the outcome of joint measurement
of Charlie and Diana is $|\psi^{\pm}\rangle$ ($|\phi^{\pm}\rangle$)
then Bob applies $U$ ($U_{1}$) on his qubits. Subsequently he measures
the auxiliary qubit. If his measurement on the auxiliary qubit yields
$|1\rangle$ then the scheme fails otherwise he recovers the unknown
state by applying appropriate unitary operations. The relation among
the measurement outcomes of Alice, outcomes of joint measurement of
Charlie and Diana, measurement outcomes of Bob, unitary operations
used by Bob and Bob's conclusions/final state is shown in the Table
\ref{tab:probabilistic-hqis-bob}. Clearly Bob needs help of both
Charlie and Diana to reconstruct the state sent by Alice as before.

\begin{table}
\begin{centering}
\begin{tabular}{|>{\centering}p{0.8in}|>{\centering}p{0.8in}|>{\centering}p{0.8in}|>{\centering}p{0.8in}|>{\centering}p{0.8in}|>{\centering}p{0.8in}|>{\centering}p{0.8in}|}
\hline 
Alice's measurement outcome  &  Outcome of joint measurement of Charlie and Diana  & Bob's two qubit operation & Bob's measurement outcome & Bob's state & Bob's operation & Final state of Bob\tabularnewline
\hline 
$|\psi^{+}\rangle$  & $|\psi^{+}\rangle_{CD}$  & $U$ & $|0\rangle$ & $\frac{|0\rangle-\lambda|1\rangle}{\sqrt{1+|\lambda|^{2}}}$ & $Z$ & $\frac{|0\rangle+\lambda|1\rangle}{\sqrt{1+|\lambda|^{2}}}$\tabularnewline
\hline 
$|\psi^{+}\rangle$  & $|\psi^{-}\rangle_{CD}$  & $U$ & $|0\rangle$ & $\frac{|0\rangle+\lambda|1\rangle}{\sqrt{1+|\lambda|^{2}}}$ & $I$ & $\frac{|0\rangle+\lambda|1\rangle}{\sqrt{1+|\lambda|^{2}}}$\tabularnewline
\hline 
$|\psi^{-}\rangle$  & $|\psi^{+}\rangle_{CD}$  & $U$ & $|0\rangle$ & $\frac{|0\rangle+\lambda|1\rangle}{\sqrt{1+|\lambda|^{2}}}$ & $I$ & $\frac{|0\rangle+\lambda|1\rangle}{\sqrt{1+|\lambda|^{2}}}$\tabularnewline
\hline 
$|\psi^{-}\rangle$  & $|\psi^{-}\rangle_{CD}$  & $U$ & $|0\rangle$ & $\frac{|0\rangle-\lambda|1\rangle}{\sqrt{1+|\lambda|^{2}}}$ & $Z$ & $\frac{|0\rangle+\lambda|1\rangle}{\sqrt{1+|\lambda|^{2}}}$\tabularnewline
\hline 
$|\phi^{+}\rangle$  & $|\psi^{+}\rangle_{CD}$  & $U$ & $|0\rangle$ & $-\frac{\left(|1\rangle-\lambda|0\rangle\right)}{\sqrt{1+|\lambda|^{2}}}$ & $XZ$ & $\frac{|0\rangle+\lambda|1\rangle}{\sqrt{1+|\lambda|^{2}}}$\tabularnewline
\hline 
$|\phi^{+}\rangle$  & $|\psi^{-}\rangle_{CD}$  & $U$ & $|0\rangle$ & $\frac{|1\rangle+\lambda|0\rangle}{\sqrt{1+|\lambda|^{2}}}$ & $X$ & $\frac{|0\rangle+\lambda|1\rangle}{\sqrt{1+|\lambda|^{2}}}$\tabularnewline
\hline 
$|\phi^{-}\rangle$  & $|\psi^{+}\rangle_{CD}$  & $U$ & $|0\rangle$ & $-\frac{\left(|1\rangle+\lambda|0\rangle\right)}{\sqrt{1+|\lambda|^{2}}}$ & $X$ & $\frac{|0\rangle+\lambda|1\rangle}{\sqrt{1+|\lambda|^{2}}}$\tabularnewline
\hline 
$|\phi^{-}\rangle$  & $|\psi^{-}\rangle_{CD}$  & $U_{1}$ & $|0\rangle$ & $\frac{|1\rangle-\lambda|0\rangle}{\sqrt{1+|\lambda|^{2}}}$ & $XZ$ & $\frac{|0\rangle+\lambda|1\rangle}{\sqrt{1+|\lambda|^{2}}}$\tabularnewline
\hline 
$|\psi^{+}\rangle$  & $|\phi^{+}\rangle_{CD}$  & $U_{1}$ & $|0\rangle$ & $\frac{|0\rangle+\lambda|1\rangle}{\sqrt{1+|\lambda|^{2}}}$ & $I$ & $\frac{|0\rangle+\lambda|1\rangle}{\sqrt{1+|\lambda|^{2}}}$\tabularnewline
\hline 
$|\psi^{+}\rangle$  & $|\phi^{-}\rangle_{CD}$  & $U_{1}$ & $|0\rangle$ & $-\frac{\left(|0\rangle-\lambda|1\rangle\right)}{\sqrt{1+|\lambda|^{2}}}$ & $Z$ & $\frac{|0\rangle+\lambda|1\rangle}{\sqrt{1+|\lambda|^{2}}}$\tabularnewline
\hline 
$|\psi^{-}\rangle$  & $|\phi^{+}\rangle_{CD}$  & $U_{1}$ & $|0\rangle$ & $\frac{|0\rangle-\lambda|1\rangle}{\sqrt{1+|\lambda|^{2}}}$ & $Z$ & $\frac{|0\rangle+\lambda|1\rangle}{\sqrt{1+|\lambda|^{2}}}$\tabularnewline
\hline 
$|\psi^{-}\rangle$  & $|\phi^{-}\rangle_{CD}$  & $U_{1}$ & $|0\rangle$ & $-\frac{\left(|0\rangle+\lambda|1\rangle\right)}{\sqrt{1+|\lambda|^{2}}}$ & $I$ & $\frac{|0\rangle+\lambda|1\rangle}{\sqrt{1+|\lambda|^{2}}}$\tabularnewline
\hline 
$|\phi^{+}\rangle$  & $|\phi^{+}\rangle_{CD}$  & $U_{1}$ & $|0\rangle$ & $\frac{|1\rangle+\lambda|0\rangle}{\sqrt{1+|\lambda|^{2}}}$ & $X$ & $\frac{|0\rangle+\lambda|1\rangle}{\sqrt{1+|\lambda|^{2}}}$\tabularnewline
\hline 
$|\phi^{+}\rangle$  & $|\phi^{-}\rangle_{CD}$  & $U_{1}$ & $|0\rangle$ & $\frac{|1\rangle-\lambda|0\rangle}{\sqrt{1+|\lambda|^{2}}}$ & $XZ$ & $\frac{|0\rangle+\lambda|1\rangle}{\sqrt{1+|\lambda|^{2}}}$\tabularnewline
\hline 
$|\phi^{-}\rangle$  & $|\phi^{+}\rangle_{CD}$  & $U_{1}$ & $|0\rangle$ & $\frac{|1\rangle-\lambda|0\rangle}{\sqrt{1+|\lambda|^{2}}}$ & $XZ$ & $\frac{|0\rangle+\lambda|1\rangle}{\sqrt{1+|\lambda|^{2}}}$\tabularnewline
\hline 
$|\phi^{-}\rangle$  & $|\phi^{-}\rangle_{CD}$  & $U_{1}$ & $|0\rangle$ & $\frac{|1\rangle+\lambda|0\rangle}{\sqrt{1+|\lambda|^{2}}}$ & $X$ & $\frac{|0\rangle+\lambda|1\rangle}{\sqrt{1+|\lambda|^{2}}}$\tabularnewline
\hline
\end{tabular}
\par\end{centering}

\caption{\label{tab:probabilistic-hqis-bob}Probabilistic HQIS (Bob reconstructs
the unknown state): The relation among the measurement outcomes of
Alice, outcomes of joint measurement of Charlie and Diana, measurement
outcomes of Bob, unitary operations used by Bob and Bob's conclusions/final
state. In the remaining 16 cases where Bob's measurement outcome is
$|1\rangle$, his final state will be $|1\rangle$ and the HQIS scheme
will fail. Global phases are ignored in the last column. }
\end{table}

Thus using this scheme Alice can hierarchically split quantum information
among her agents and one of the agents can recover the unknown quantum
state with unit fidelity. However, the success rate of the scheme
is not unity. This is why it can be referred to as probabilistic HQIS
in analogy with the conventional probabilistic teleportation scheme.

\section{Hierarchical quantum secret sharing (HQSS)\label{sec:HQSS}}

The QSS scheme was originally proposed in 1999 \cite{Hillery}. The
scheme may now be generalized and modified as follows. As before consider
that Alice is boss of a company and Bob, Charlie and Diana are her
agents. Further, consider that Alice lives in Amsterdam. Bob, Charlie
and Diana are her agents in Berlin. Alice wants to send them a secret
message to perform a job. However, one of them may be dishonest and
Alice does not know who is dishonest. But Diana is a senior employee
of the company and she has been working in the company since long,
while Bob and Charlie has joined recently. So Alice trusts Diana more
than the other two agents. Thus there is a hierarchy among the agents.
In this situation, Alice may use HQIS scheme with 4-qubit $|\Omega\rangle$
as described in Subsection \ref{sub:Case-I:-} and send the information
in three pieces so that none of Bob, Charlie and Diana can read the
message of Alice without the help of the others. However, Diana would
require lesser help than Bob. Now there exist  possibilities of eavesdropping.
For example, consider that Bob is dishonest and he captures the qubit
sent to Charlie and Diana too. If Bob does a Bell measurement on Charlie's
and Diana's qubit then using the unitary operations described in Table
\ref{tab:Table3}, he will be able to get the entire information without
any help of Charlie and Diana. Consequently, to circumvent this kind
of attack from a dishonest user or an external eavesdropper, Alice
needs to add some error checking schemes to the above proposed HQIS
schemes. One way to achieve this is as follows:
\begin{description}
\item [{HQSS~1:}] Alice prepares $|\Omega\rangle^{\otimes n}$ . As $|\Omega\rangle^{\otimes n}$
is a $4n$-qubit state, qubits of $|\Omega\rangle^{\otimes n}$ may
be indexed as $p_{1},p_{2},\cdots,p_{4n}$. Thus $p_{s}$ is the $s^{{\rm th}}$
qubit of $|\Omega\rangle^{\otimes n}$ and $\{p_{4l-3},p_{4l-2},p_{4l-1},p_{4l}:l\leq n\}$
are the $4$ qubits of the $l^{{\rm th}}$ copy of $|\Omega\rangle^{\otimes n}$.
\\

\item [{HQSS~2:}] Using all the first qubits of her possession, Alice
creates an ordered sequence $P_{A}=\left[p_{1},p_{5},p_{9},\cdots,p_{4n-3}\right]$.
Similarly she prepares an ordered sequence with all the second qubits
as $P_{B}=\left[p_{2},p_{6},p_{10},\cdots,p_{4n-2}\right]$, another
ordered sequence with all the third qubits as $P_{C}=\left[p_{3},p_{7},p_{11},\cdots,p_{4n-1}\right]$
and another ordered sequence with all the fourth qubits as $P_{D}=\left[p_{4},p_{8},p_{12},\cdots,p_{4n}\right].$
She prepares $3n$ decoy qubits\textcolor{red}{{} }$d_{i}$ with $i=1,2,\cdots,3n$
such that $d_{i}\in\{|0\rangle,|1\rangle,|+\rangle,|-\rangle\}$\textcolor{red}{{}
}and concatenates first $n$ of them with $P_{B}$ to yield a larger
sequence $P_{B^{\prime}}=\left[p_{2},p_{6},p_{10},\cdots,p_{4n-2},d_{1},d_{2},\cdots,d_{n}\right]$.
Similarly, using $P_{C}$ and next $n$ decoy qubits she creates $P_{C^{\prime}}=\left[p_{3},p_{7},p_{11},\cdots,p_{4n-1},d_{n+1},d_{n+2},\cdots,d_{2n}\right]$
and using $P_{D}$ and the last $n$ decoy qubits she creates $P_{D^{\prime}}=\left[p_{4},p_{8},p_{12},\cdots,p_{4n},d_{2n+1},d_{2n+2},\cdots,d_{3n}\right]$.
Thereafter Alice applies a permutation operator $\Pi_{2n}$ on $P_{B^{\prime}},$
$P_{C^{\prime}}$ and $P_{D^{\prime}}$ to create random sequences
$P_{B^{\prime\prime}}=\Pi_{2n}P_{B^{\prime}}$, $P_{C^{\prime\prime}}=\Pi_{2n}P_{C^{\prime}}$
and $P_{D^{\prime\prime}}=\Pi_{2n}P_{D^{\prime}}$ and sends $P_{B^{\prime\prime}}$,
$P_{C^{\prime\prime}}$ and $P_{D^{\prime\prime}}$ to Bob, Charlie
and Diana respectively. The actual order is known to Alice only. 
\item [{HQSS~3:}] Alice announces $\Pi_{n}\in\Pi_{2n}$, the coordinates
of the decoy qubits in each sequence after receiving authenticated
acknowledgments of receipt of all the qubits form Bob, Charlie and
Diana. The BB84 subroutine %
\footnote{BB84 subroutine means eavesdropping is checked using conjugate coding
in a manner similar to what followed in BB84 protocol. Explicitly,
in our particular case, Alice's announcement of the position of the
decoy qubits provides a verification string to each of her agents.
Now a agent measures either all the qubits of the verification string
randomly in $\left\{ 0,1\right\} $ or $\left\{ +,-\right\} $ basis
and announces which basis she (he) has used to measure a particular
qubit, position of that qubit in the string and outcome. Alice compares
the initial states of the decoy qubits with the outcomes of an agent
in all those cases where the basis used by her to prepare the decoy
qubit is same as the basis used by the particular agent to measure
it. Ideally in absence of eavesdropping it should match. Any eavesdropping
effort would lead to mismatch.%
} to detect eavesdropping, is then implemented on the decoy qubits
by measuring them in the nonorthogonal bases $\left\{ |0\rangle,|1\rangle\right\} $
or $\left\{ |+\rangle,|-\rangle\right\} $. If sufficiently few errors
are found in all the sequences, then they go to the next step; else,
they return to \textbf{HQSS~1}. \\
This will ensure that the initial $|\Omega\rangle^{\otimes n}$
state is appropriately distributed among Alice, Bob, Charlie and Diana
without any eavesdropping. As the eavesdropping is checked by the
BB84 subroutine, security of the protocol is equivalent to that of
BB84 protocol.
\item [{HQSS~4:}] Alice discloses the coordinates of the remaining qubits
and Bob, Charlie and Diana rearrange their sequences accordingly.\\
Remaining part of the protocol is same as HQIS scheme described
in Subsection \ref{sub:Case-I:-}. Now consider that Alice's quantum
secret which is to be shared is $|\psi_{s}\rangle=\frac{1}{\sqrt{1+|\lambda|^{2}}}(|0\rangle+\lambda|1\rangle)$.
Thus Alice, Bob, Charlie and Diana share $n$ 5-qubit states of the
form (\ref{eq:chitra2}), where first two qubits are with Alice and
the last three qubits are with Bob, Charlie and Diana, and $|\psi_{0}\rangle_{BCD}=\frac{1}{\sqrt{2}}[|000\rangle+|110\rangle]$
and $|\psi_{1}\rangle_{BCD}=\frac{1}{\sqrt{2}}[|001\rangle-|111\rangle]$. 
\item [{HQSS~5:}] Alice measures her qubits in Bell basis and announces
the results.\\
Without loss of generalization we may assume that Alice has asked
Bob to prepare the secret state transmitted by her.
\item [{HQSS~6:}] Charlie and Diana jointly measures their qubits using
Bell basis and communicate the result to Bob.
\item [{HQSS~7:}] Bob applies appropriate unitary operators (as described
in Table \ref{tab:Table3}) in accordance with the measurement outcomes
of Alice, Charlie and Diana, and reconstructs the secret quantum state
transmitted by Alice.\\

\end{description}
If Alice asked Diana to prepare the secret state transmitted by her,
then the last two steps should be modified as follows:
\begin{description}
\item [{HQSS~6:}] Bob (Charlie) measures his qubit using computational
basis and communicates the result to Diana.
\item [{HQSS~7:}] Diana applies appropriate unitary operators (as described
in Table \ref{tab:Table2}) in accordance with the measurement outcomes
of Alice and Bob (Charlie), and reconstructs the secret quantum state
transmitted by Alice.
\end{description}
This is clearly a scheme of HQSS as Diana requires lesser information
than Bob to obtain the secret shared by Alice. This provides a clear
and unconditionally secure solution to the practical problem described
in the beginning of this section where Alice wishes to distribute
a quantum state among her agents in an asymmetric and secure manner.
It also provides a solution to the practically relevant problem related
to the use of nuclear weapon described in Section \ref{sec:Introduction}.
Here it would be apt to note that we have explicitly described a scheme
for HQSS using $|\Omega\rangle$ state. However, the protocol described
above is much more general and any quantum state where HQIS is possible
may be used to implement HQSS. For example, 4-qubit cluster state,
6-qubit cluster state and graph state can be used to implement HQSS.
Further, it is now a straightforward task to turn this HQSS scheme
into a probabilistic HQSS scheme. In the earlier studies of Wang \emph{et
al.} possibility of HQSS was hinted but neither an explicit protocol
was provided nor the security threats arising from one of the user
being dishonest was properly discussed. Thus the present scheme is
the first ever scheme of HQSS which is expected to find applications
in many important practical situations.

\section{Conclusions\label{sec:Conclusions}}

We have generalized Wang \emph{et al.}'s idea of asymmetric quantum
information splitting from various perspectives. To be precise, we
have provided a more general framework to study the hierarchical quantum
information splitting (HQIS) and have shown that the same can be modified
to yield protocols of HQSS and probabilistic HQIS. The generalization
is important for several reasons. Firstly, there exist several practical
situations where the asymmetric information splitting (Especially
HQSS) is relevant. Secondly, the possibility of HQSS and probabilistic
HQIS were not investigated earlier. Following a similar line of arguments
we can also obtain a scheme for hierarchical joint remote state preparation.
That would be shown elsewhere \cite{HRSP}. Further the approach adopted
in the present paper can be easily used to explore the possibilities
of observing HQIS, probabilistic HQIS and HQSS in other quantum states. 

\textbf{Acknowledgment:} A. P. thanks Department of Science and Technology
(DST), India for support provided through the DST project No. SR/S2/LOP-0012/2010.
He also thanks the Operational Program Education for Competitiveness
- European Social Fund project CZ.1.07/2.3.00/20.0017 of the Ministry
of Education, Youth and Sports of the Czech Republic. Authors also
thank Dr. R. Srikanth for some helpful technical discussions.


\begin{thebibliography}{16}
\bibitem{Bennet1993}C.H. Bennett. G. Brassard, C. Crépeau, R. Jozsa,
A. Peres and W. K. Wootters, Phys. Rev. Lett., \textbf{70 }(1993)
1895.

\bibitem{ijqi-qis}A. Pathak and A. Banerjee, , Int. J. Quan. Info.
\textbf{9} (2011) 389.

\bibitem{probalistic-1}W.-L. Li, C.-F. Li and G.-C Guo, Phys. Rev.
A \textbf{61} (2000) 034301. 

\bibitem{probabilistic-2}Z.-L. Cao, M. Yang, G.-C. Guo, Phys. Lett.
A \textbf{308 }(2003) 349. 

\bibitem{probabilistic-3}H. Lu, G.-C. Guo, Phys. Lett. A \textbf{276}
(2000) 209. 

\bibitem{probabilistic-4}B.-S. Shi, Y.-K. Jiang, G.-C. Guo, Phys.
Lett. A \textbf{268} (2000) 161.

\bibitem{qis-panigrahi} S. Muralidharan and P. K. Panigrahi, Phys.
Rev. A \textbf{78} (2008) 062333.

\bibitem{pankaj-qis-W}P. Agrawal and A. Pati, Phys. Rev. \textbf{74}
(2006) 062320.

\bibitem{opt. comm}X.-W. Wang, L.-X. Xiac, Z.-Y. Wangd, D.-Y. Zhang,
Opt. Comm. \textbf{283} (2010) 1196.

\bibitem{jphysb}X.-W. Wang, D.-Y. Zhang, S.-Q. Tang and L.-J. Xie,
J. Phys. B \textbf{44} (2011) 035505.

\bibitem{ijtp}X.-W. Wang, D.-Y. Zhang, S.-Q. Tang, X.-G. Zhan, K.-M.You,
Int. J. Theor. Phys. \textbf{49} (2010) 2691. 

\bibitem{Hillery}M. Hillery, V. Buzek and A. Bertaiume, Phys. Rev.
A \textbf{59} (1999) 1829.

\bibitem{Pati-RSP}A.K. Pati, Phys. Rev. A \textbf{63} (2000) 014302.

\bibitem{Ba An -RSP}N.B. An and J. Kim, J. Phys. B. \textbf{41} (2008)
095501.

\bibitem{Pradhan-Agarwal-Pati}B. Pradhan, P. Agrawal and A.K. Pati,
arXiv:0705.1917v1 (quant-ph).

\bibitem{HRSP}C. Shukla and A. Pathak, Hierarchial joint remote state
preparation, under preparation.
\end{thebibliography}
\end{document}